\documentclass[twocolumn,showpacs,showkeys,superscriptaddress]{revtex4}
\usepackage{amsmath}
\usepackage{amsfonts}
\usepackage{amssymb}
\usepackage{pbsi}
\usepackage[T1]{fontenc}
\usepackage{hyperref}
\usepackage{xcolor}

\def\openone{\leavevmode\hbox{\small1\kern-3.8pt\normalsize1}}
\def\N{\leavevmode\hbox{ Z \kern-8 pt\normalsize{Z}}}
\def\openone{\leavevmode\hbox{\small1\kern-3.8pt\normalsize1}}
\def\openJ{\leavevmode\hbox{J \kern-9.5pt\normalsize J}}
\def\openS{\leavevmode\hbox{ S \kern-9.3pt\normalsize S}}
\newcommand{\bb}{\begin{equation}}
\newcommand{\ee}{\end{equation}}
\newcommand{\eqb}{\begin{eqnarray}}
\newcommand{\eqf}{\end{eqnarray}}

\usepackage{color}

\begin{document}

\title{A new approach to solving the Schr\"odinger equation}

\author{Sergio A. Hojman}
\email{sergio.hojman@uai.cl}
\affiliation{Departamento de Ciencias, Facultad de Artes Liberales,
Universidad Adolfo Ib\'a\~nez, Santiago 7491169, Chile.}
\affiliation{Centro de Investigaci\'on en Matem\'aticas, A.C., Unidad M\'erida, Yuc. 97302, M\'exico}
\affiliation{Departamento de F\'{\i}sica, Facultad de Ciencias, Universidad de Chile,
Santiago 7800003, Chile.}
\affiliation{Centro de Recursos Educativos Avanzados,
CREA, Santiago 7500018, Chile.}
\author{Felipe A. Asenjo}
\email{felipe.asenjo@uai.cl}
\affiliation{Facultad de Ingenier\'ia y Ciencias,
Universidad Adolfo Ib\'a\~nez, Santiago 7491169, Chile.}

\begin{abstract}
A new approach to find exact solutions to one--dimensional quantum mechanical systems is devised. The scheme is based on the introduction of a potential function for the wavefunction, and the equation it satisfies. We recover known solutions as well as to get new ones for both free and interacting particles with wavefunctions having vanishing and non--vanishing Bohm potentials. For most of the potentials, no solutions to the Schr\"odinger equation produce a vanishing Bohm potential. A (large but) restricted family of potentials allows the existence of particular solutions for which the Bohm potential vanishes. This family of potentials is determined, and several examples are presented. It is shown that some quantum, such as accelerated Airy wavefunctions, are due to the presence of non--vanishing Bohm potentials. New examples of this kind are found and discussed.  
\end{abstract}

%\pacs{04.60 Bc, 98.80 Qc, 11.30 Cp}

\maketitle
\section{Introduction}

The Madelung--Bohm approach to Schr\"odinger's equation  \cite{mad,bohm}, written in terms of the amplitude and the phase of the quantum wavefunction, gives rise to two real equations. The first one is a Quantum Hamilton--Jacobi equation (QHJE) that is similar to its classical counterpart. It differs from it by the addition of an extra term, the Bohm potential, which depends on the amplitude of the wavefunction only. The second one is the continuity (probability conservation) equation \cite{mad,bohm,holland,wyatt,ha20}. 

The presence of the Bohm potential in the QHJE gives rise to unexpected solutions for the wavefunction that, in turn, behave in surprising ways that seem to be at odds with their classical counterparts. One of the most striking results, accelerating ``free'' quantum particles, was predicted by Berry and Balazs in 1979 \cite{berry} and found experimentally, in an optical setting, by Siviloglou, Broky, Dogariu, and Christodoulides in 2007 \cite{sivi}. The best known free quantum particle wavefunctions have amplitudes whose Bohm potential vanishes and, therefore, their QHJEs are identical to the classical Hamilton--Jacobi equations and they present no surprises. On the other hand, the Berry--Balazs solution exhibits non--vanishing Bohm potential and a departure from its classical limit.
This kind of unexpected behavior due to the presence of the Bohm potential appears in any wave equation irrespective of its classical or quantum origin, as it does in the scalar wave equation, in Maxwell's equations, in gravitational waves equations \cite{ha20}.

The main aim of this work is to propose a mechanism to find exact solutions to the Schr\"odinger equation where we can explicitly identify the contribution of Bohm potential. In that way, we expect to obtain new solutions, with interesting features, for free particles as well as for systems with non--trivial potentials. To contribute deepening the understanding of the role of Bohm potential in Quantum Mechanics, we classify potentials according to whether or not they admit solutions with vanishing Bohm potentials for one-dimensional systems. It is shown that the system can be completely solved by re--writing the wavefunction in terms of a potential field ${{f}}$. A non--linear integro--differential equation is derived for ${{f}}$. This  equation, as far as we know, is new and it is equivalent to  Schr\"odinger's. It provides new insights in the interrelation between the amplitude and the phase of the wavefunction.
In general, we solve exactly the Schr\"odinger equation for families of solutions with non--constant Bohm potentials for a wide variety of external potentials \cite{Makowski}. Finally, we also discuss the pertinence of these results to the applicability of the Van Vleck--Morette determinant approximation \cite{vanvleck,morette}.

Other authors have also explored alternative ways to deal with the Schr\"odinger equation by studying Hamiltonians which satisfy different kinds of particular conditions. See, for instance, Refs.~\cite{berry1,demir1,demir2,berry2,guery}.

\section{General approach in one--dimensional quantum mechanics}

For a real potential $V({ x},t)$, let us consider the Schr\"odinger equation in a one--dimensional space
\begin{equation}
\left[-\frac{{\hbar}^2}{2m}\frac{\partial^2}{\partial x^2} + V ({x},t) - i \hbar \frac{\partial}{\partial t}  \right] \psi ({ x},t) = 0, \label{schr}
\end{equation}
for wavefunction $\psi=\psi({ x},t)$, and its complex conjugate counterpart.
Therefore, a solution for the wavefunction in terms of the polar form can be invoked \cite{holland,wyatt}
\begin{equation}\label{psi}
{{\psi}}({ x},t) = A({ x},t)\ e^{i S({x},t)/\hbar}\, ,     \end{equation}
where now $A({ x},t)$ and $S({ x},t)$ are real functions of  space and of time. This solution produces that the real and imaginary parts of Schr\"odinger equations become, respectively,
\begin{eqnarray}
\frac{1}{2m} S'^2 - \frac{\hbar^2}{2m}\frac{  A'' }{A} + V +\dot S&=&0\, , \label{HJB1} \\
\frac{1}{m}\left(A^2 S'\right)'  +\dot A^2 &=&0\, . \label{cont}    
\end{eqnarray}
where  and
$S'\equiv\partial_x S$, and $\dot S\equiv \partial_t S$ (and similarly for $A$).
The first equation \eqref{HJB1} is a modified version of the classical Hamilton--Jacobi equation for the potential $V$ \cite{holland,wyatt}, while the second one \eqref{cont} is simply the continuity (probability conservation) equation. The classical Hamilton--Jacobi equation is modified by the addition of the Bohm potential $V_B$ in one dimension
\begin{equation}
V_B \equiv -\frac{\hbar^2}{2m}\frac{ A'' }{A}\, , \label{VB2}    
\end{equation}
which can be interpreted as an internal energy \cite{glen}.

Rather than of solving the previous system, let us introduce the function ${f}={ f}({ x},t)$, defined by the following relations
\begin{eqnarray}
A({ x},t)&=&\sqrt{f'} \label{A}\, ,\nonumber\\
S({ x},t)&=&\mu(t)-m\int \frac{\,  \dot{ f}}{f'}dx \label{S}\, .
\end{eqnarray}
for an arbitrary time--dependent function $\mu(t)$.
The introduction of the potential field ${ f}$ transforms the continuity equation \eqref{cont} in an identity.
However, more important, it is the fact that knowing ${ f}$, the fields $A$ and $S$ are found, and  thus the wavefunction is completely determined.

In order to find the evolution of field ${ f}$, we need to evaluate a
modified version of the classical Hamilton--Jacobi equation \eqref{HJB1}. This equation is now re-written as
\begin{eqnarray}
&&\frac{1}{2} \frac{m {\dot f}^2}{f'^2} - \frac{\hbar^2}{4m}\left(\frac{f'''}{f'}-\frac{1}{2}\frac{f''^2}{f'^2} \right) + \nonumber\\
&&\qquad m \int_{\tilde x=0}^{\tilde x=x} \left(\frac{\dot f \dot f'}{f'^2}-\frac{\ddot f}{f'} \right) d\tilde x +\dot\mu+V =0, \label{newschV}
\end{eqnarray}
For a given potential $V$, this equation  establishes the evolution of field ${ f}$.
However, this equation can also be read in the opposite way. Starting from any ${ f}$, the equation defines the potential that solves Schr\"odinger equation. Also, the gradient of Eq.~\eqref{newschV} can be considered an equation for the force ${ F}= - V'$,
 by differentiating Eq.~\eqref{newschV}  with respect to $x$,
\begin{eqnarray}
&&\left(\frac{1}{2} \frac{m {\dot f}^2}{f'^2} - \frac{\hbar^2}{4m}\left(\frac{f'''}{f'}-\frac{1}{2}\frac{f''^2}{f'^2} \right)\right)'-\nonumber \\ &&F+m  \left(\frac{\dot f \dot f'}{f'^2}-\frac{\ddot f}{f'} \right) =0\, . \label{newschF}
\end{eqnarray}

It is important to remark that the above solutions for $A$ and $S$ are general for a one--dimensional system. This implies that they provide a general solution for the amplitude which is, in general, different from the Van Vleck--Morette expression (VVM) \cite{vanvleck,morette} for one--dimensional systems
\begin{equation}
    A=\sqrt{\frac{1}{2 i \pi \hbar}\frac{\partial^2 S}{\partial x_f\partial x_i}}\, ,
    \label{condVVM}
\end{equation}
with initital $x_i$ and final $x_f$ positions.
The VVM result represents an approximation to the amplitude (in the WKB approach sense) of the Schr\"odinger equation wavefunction. Below we show explicit examples for exact solutions \eqref{newschV} which coincide with the approximate VVM expression \eqref{condVVM}. 
On the other hand, we show other examples of exact solutions that do not coincide with the VVM expression for the amplitude of the wave function.

It is clear that Bohm potential \eqref{VB2} plays a remarkable role in the solutions one can find for a given potential $V$. The Bohm potential is the origin to the difference between classical and quantum dispersion relations that emerge from the modified version of the classical Hamilton--Jacobi equation \cite{ha20}. For this reason, it is worth studying Eq.~\eqref{newschV} for free and interacting particles classifying the solutions according to whether or not they produce a vanishing Bohm potential \eqref{VB2}. We explore different potentials $V$, for each case.
We prove below that the condition of producing a vanishing Bohm potential gives rise to a general family of time--dependent forces.

\section{Vanishing Bohm potential}

The WKB approximation in Quantum Mechanics and the eikonal approach which leads to Geometrical Optics deal with solutions which yield a negligible Bohm potential \cite{berry72}, i.e., a Bohm potential which is approximately equal to zero (either because $\hbar$ or the wavelength of light are considered small, in the quantum mechanical or optical cases, or the amplitude is assumed to vary slowly in either case).
In this section we will find the most general one--dimensional potential $V$ which admits solutions which give rise to a Bohm potential \eqref{VB2} which is exactly equal to zero.

For a vanishing Bohm potential \eqref{VB2}, the function $f$ must satisfy  
\begin{equation}
\frac{A''}{A} \equiv   \frac{1}{2}\left(\frac{f'''}{f'}-\frac{1}{2}\frac{f''^2}{f'^2} \right)=0\ . \label{W} 
\end{equation}
The general solution for function $f(x,t)$ reads
\begin{equation}
f(x,t)=\frac{a(t)^2}{3} x^3+a(t)b(t) x^2+b(t)^2 x+c(t), \label{fW}   
\end{equation}
for arbitrary functions $a(t)$, $b(t)$ and $c(t)$. In this case, Eq.~\eqref{newschV} reduces to
\begin{eqnarray}
\frac{1}{2} \frac{m {\dot f}^2}{f'^2} + m \int_{\tilde x=0}^{\tilde x=x} \left(\frac{\dot f \dot f'}{f'^2}-\frac{\ddot f}{f'} \right) d\tilde x +\dot\mu+V =0, \label{newschV2}
\end{eqnarray}
The force  ${F}=-V'$, derived from the above potential, for vanishing Bohm poential, is readily obtained as
\begin{eqnarray}
F(x,t)&=& \frac{m}{2}\left( \frac{{\dot f}^2}{f'^2}\right)'  +m \left( \frac{\dot f \dot f'}{f'^2}-\frac{\ddot f}{f'} \right)\nonumber\\  
&=&-\frac{1}{
9(x a(t)+b(t))^5}\cdot \nonumber \\ &&
[m x^6 \left(-10 a(t)^3 \dot a(t)^2+6 a(t)^4 \ddot a(t)\right)+ \nonumber \\
&&mx^5 (-42 a(t)^2 b(t) \dot a(t)^2-18 a(t)^3 \dot a(t) \dot b(t)+\nonumber \\ &&27 a(t)^3 b(t) \ddot a(t)+9 a(t)^4 \ddot b(t))+\nonumber \\  
&&mx^4 (-60 a(t) b(t)^2 \dot a(t)^2-90 a(t)^2 b(t) \dot a(t) \dot b(t)+\nonumber \\ && 45 a(t)^2 b(t)^2 \ddot a(t)+45 a(t)^3 b(t) \ddot b(t))+\nonumber \\ 
&&mx^3 (-30 b(t)^3 \dot a(t)^2-150 a(t) b(t)^2 \dot a(t) \dot b(t)-\nonumber \\ 
&&18 a(t)^2 b(t) \dot b(t)^2-12 a(t)^2 \dot a(t) \dot c(t)+\nonumber \\ &&33 a(t) b(t)^3 \ddot a(t)+81 a(t)^2 b(t)^2\ddot b(t)+\nonumber \\
&&9 a(t)^3 \ddot c(t))+\nonumber \\ 
&&mx^2 (-90 b(t)^3 \dot a(t) \dot b(t)-54 a(t) b(t)^2 \dot b(t)^2-\nonumber \\ 
&& 36 a(t) b(t) \dot a(t) \dot c(t)+9 b(t)^4 \ddot a(t)+\nonumber \\ &&63 a(t) b(t)^3 \ddot b(t)+27 a(t)2 b(t) \ddot c(t))+\nonumber \\
&&mx (-54 b(t)^3 \dot b(t)^2-36 b(t)^2 \dot a(t) \dot c(t)+\nonumber \\ &&18 b(t)^4 \ddot b(t)+27 a(t) b(t)^2 \ddot c(t))+\nonumber \\
&&m(-36 b(t)^2 \dot b(t) \dot c(t)+18 a(t) \dot c(t)^2+\nonumber \\ &&9 b(t)^3 \ddot c(t))]\ .\label{FW}
\label{newschFW}
\end{eqnarray}
This force (or its associated one--dimensional potential $V$) is the most general one whose Schr\"odinger equations admit solutions with vanishing Bohm potential, i.e., such that $V_B(x,t)=0$.
This depends on the different and arbitrary choices of the time--dependent functions $a(t)$, $b(t)$ and $c(t)$. 
On the other hand, every time that $f$ does not have the form \eqref{fW}, Bohm potential does not vanish.

Below we study different exact solution that exemplify those cases.

\section{Free particles with vanishing Bohm potential}

Several solutions for the quantum free particle, with $V=0$, produce a vanishing Bohm potential.

\subsection{Free particle as a plane wave}
\label{secplanewave}

Using  \eqref{fW}, let us choose
\begin{eqnarray}
a(t)=0\, ,\qquad b(t)=1\, ,\qquad c(t)=-\frac{k}{m} t\, , \end{eqnarray}
where  $k$ is a constant with units of inverse length. This functions allow us to study a free particle with vanishing Bohm potential. This produces the amplitude and phase
\begin{eqnarray}
A(x,t)&=& 1\, ,\nonumber\\
S(x,t)&=& k x-\frac{k^2}{2m} t\, ,
\label{freeplanewave}
\end{eqnarray}
where $\mu(t)=k^2 t/2m$.
This quantum solution represents a free particle as a plane wave, with a wave phase velocity equal to $k/2m$.

\subsection{Non--separable solution for the quantum free particle}
\label{soludiffusion1}

The above solution can be easily found by traditional approaches by separating the functionality of space and time in the wavefunction. However, there are other solutions for quantum free particles, where this is not possible.

Let us choose $f$ from \eqref{fW}, such that
\begin{eqnarray}
a(t)=\sqrt{\frac{\alpha}{\left(t-t_i\right)^{3}}}\, ,\quad b(t)=\sqrt{\frac{\beta}{t-t_i}}\, ,\quad c(t)=\gamma\, , \end{eqnarray}
with constant $\alpha$, $\beta$, $\gamma$, and initial time $t_i$. Then, the amplitude and phase \eqref{S} becomes
\begin{eqnarray}
A(x,t)&=& \frac{\sqrt{\alpha}\,  x}{\left(t-t_i\right)^{3/2}}+\sqrt{\frac{\beta}{t-t_i}} \, ,\nonumber\\
S(x,t)&=& \frac{m\, x^2}{2 \left(t - t_i\right)}\, ,
\end{eqnarray}
with $\mu=0$.
This is a solution for the free particle, with $V=0$.

\section{Free particles with  non--vanishing Bohm potential}

The previous cases are known free particle solutions. What is probably not as well--known 
is that they give rise to a vanishing Bohm potential. This opens the possibility to look for other free particle solutions with $V=0$ and a non--vanishing Bohm potential. 

In this section we show some of them for time--dependent amplitudes. In these cases, the function $f$ does not satisfy \eqref{fW}.
This is  an important fact since the non--vanishing Bohm potential introduces unexpected effects such as self--acceleration in exact solutions, as we show below.

\subsection{Free particle as a non--plane wave with arbitrary velocity}

We show in Sec.~\ref{secplanewave} that a free particle as a plane wave must have a vanishing Bohm potential.
A non--plane wave has a variable amplitude, and then, produces Bohm potential different from zero.

For the current case, let us consider the function
\begin{eqnarray}
f(x,t)=\frac{1}{\lambda}\exp\left(\lambda\, x-\frac{\hbar\, \lambda\,  k}{m}t \right)\, ,
\end{eqnarray}
where $\lambda$, and $k$ are arbitrary constants with units of inverse length. This $f$ produce a wavefunction that solves the free particle problem with $V=0$.

In this case, we can generate the amplitude
and phase of the wavefunction
\begin{eqnarray}
A(x,t)&=& \sqrt{ \exp\left(\lambda\, x-\frac{\hbar\, \lambda\,  k}{m}t \right)}\, ,\nonumber\\
S(x,t)&=&\hbar k x+\mu(t)\, ,\nonumber\\
\mu(t)&=&-\frac{\hbar^2 k^2 t}{2m}\left(1-\frac{\lambda^2}{4k^2} \right) \, .
\end{eqnarray}
This wave is different to the plane wave \eqref{freeplanewave}, as it has a constant Bohm potential \eqref{VB2}, given by
\begin{equation}
    V_B=-\frac{\hbar^2\lambda^2}{8m}\, .
\end{equation}
A vanishing Bohm potential  is obtained only in the case $\lambda=0$, thus recovering a plane wave solution.
 On the other hand, the phase velocity of this wave is 
\begin{equation}
    \frac{\hbar k}{2m}\left(1-\frac{\lambda^2}{4k^2} \right)
\end{equation}
which can be as small as it is required, when $\lambda\rightarrow 2 k$. 
However, this solution has a constant velocity, as no force (and thus no acceleration) is applied on the evolution of the particle. This can also be understood as, for this case, $V_B'=0$.

\subsection{Accelerating Airy wave packets}
\label{airysectionberry}

It is interesting to show how the very well--known Airy wave packets \cite{berry, sivi, besiers, Esat,Greenberger,kaminer,Matulis} is obtained in our formalism. We prove below that the acceleration experienced by these wave packets is due to the Bohm potential.

Let us consider 
\begin{equation}
    f(x,t)=\int \left[{\mbox Ai} \left(\frac{\beta}{\hbar^{2/3}} \left(x-\frac{\beta^3}{4m^2} t^2 \right)\right)\right]^2 dx\, ,
\end{equation}
with a non--zero constant $\beta$. For this case
\begin{eqnarray}
A(x,t)&=&{\mbox Ai} \left(\frac{\beta}{\hbar^{2/3}} \left(x-\frac{\beta^3}{4m^2} t^2 \right)\right)\, ,\nonumber\\
S(x,t)&=&\frac{\beta^3 t}{2m}\left(x - \frac{\beta^3}{6m^2} t^2\right)\, ,
\end{eqnarray}
with $\mu(t)=-\beta^6 t^2/(12 m^3)$. This solution for free particles, with $V=0$, has a non--vanishing time-- and space--dependent Bohm potential given by
\begin{equation}
    V_B(x,t)=-\frac{\beta^3}{2m}\left(x-\frac{\beta^3}{4m^2}t^2\right)\, .
\end{equation}
The Bohm potential is responsible for the constant acceleration $a_{Airy}$ experienced by the Airy wave packet 
\begin{equation}
   a_{Airy} = -\frac{V_B'}{m}=\frac{\beta^3}{2m^2}\, .
    \label{Bohmairy}
\end{equation}
Note that this result is consistent with the velocity of the Airy package $v_{Airy} = p_{Airy}/m$ where the momentum $p_{Airy}$ is given by $p_{Airy}=\partial S(x,t)/ \partial x$.
Therefore, there is a solution to the free Schr\"odinger equation which has a constant acceleration given by \eqref{Bohmairy} in spite of being in the presence of a vanishing (external) force.

This solution represents a wavepacket constructed by a non--normalizable superposition of constant velocity plane wave solutions for free particles, but a normalizable solution can be constructed from it \cite{lekner}, which can be shown to travel without acceleration.

\subsection{A different solution for a free particle}

We can find a different set of solutions for $V=0$ in the following way. Consider 
\begin{equation}
    f(x,t)=\int \left[\frac{1}{\sqrt{\sigma(t)}} Z\left(\frac{x}{\sigma(t)}\right)\right]^2 dx\, ,
\end{equation}
where $\sigma(t)$ is an arbitrary time--dependent function, and $Z$ is also an arbitrary function of argument $x/\sigma(t)$. This function generates the following one--dimensional amplitude and phase
\begin{eqnarray}
A(x,t)&=&\frac{1}{\sqrt{\sigma(t)}} Z\left(\frac{x}{\sigma(t)}\right)\, ,\nonumber\\
S(x,t)&=&\frac{m\, \dot\sigma(t)}{2\, \sigma(t)}x^2+\mu(t)\, .
\label{phaseGenSdsafaga}
\end{eqnarray}

For this case, Eq. \eqref{fW} becomes
\begin{eqnarray}
\frac{m\, \ddot\sigma}{2\, \sigma}x^2+\dot\mu=\frac{\hbar^2}{2m\sigma^2 Z}\frac{d^2 Z}{d y^2}=-V_B\, ,
\label{eqgenefreepartVb}
\end{eqnarray}
with derivatives respect to the argument $y\equiv x/\sigma$. 
The only form to solve Eq.~\eqref{eqgenefreepartVb}
in a general fashion is when
\begin{equation}
    \frac{1}{Z}\frac{d^2 Z}{d y^2}=\zeta_1\,  \frac{x^2}{\sigma^2}+\zeta_2\, .
    \label{equacionparaZ}
\end{equation}
for arbitrary constants $\zeta_1$ and $\zeta_2$.
In such case, the functions $\sigma$ and $\mu$ are completely determined by the equations
\begin{eqnarray}
\sigma&=&\sqrt{\alpha t^2+\beta t+\frac{\beta^2}{4\alpha}+\frac{\hbar^2\zeta_1}{\alpha m^2}}\, ,\nonumber\\
\mu&=&\frac{\hbar \zeta_2}{4\sqrt{\zeta_1}}\arctan\left(\frac{m}{2\hbar\sqrt{\zeta_1}} \left(2\alpha t+\beta \right)\right)\, ,
\end{eqnarray}
for constants $\alpha$ and $\beta$. Thereby, the problem of a free particle is solved. Notice that Eq.~\eqref{equacionparaZ} implies that Bohm potential \eqref{eqgenefreepartVb} is non--zero in general.
With these results, phase \eqref{phaseGenSdsafaga}
results to be a generalization of phase of Sec.~\ref{soludiffusion1}.

Any function $Z$
satisfying \eqref{equacionparaZ} produces solutions. One  example is a Gaussian wavepacket
\begin{eqnarray}
Z=\exp\left(-q\frac{x^2}{\sigma^2}\right)\, ,
\label{phaseGenSdsafaga2}
\end{eqnarray}
for some constant $q$.
that gives $\zeta_1=4q^2$, and $\zeta_2=-2q$. These solutions are for free particles, and they do not hold, for example, if the quantum system has diffusion \cite{tsekov}.

Other example is the families of  functions that produces $\zeta_1=0$ and $\zeta_2={\mbox{constant}}$. Some of those funcions are $\sin$, $\cos$, $\sinh^{-1}$, $\cosh^{-1}$, among others.

The most general functions satisfying \eqref{equacionparaZ} are Weber functions, also called Parabolic cylinder functions \cite{zweill}. Once, the complete solution for the Parabolic cylinder functions
is proposed for arbitrary $\zeta_1$ and $\zeta_2$, the whole problem is solved.

\section{Particles under non--zero external potential with vanishing Bohm potential}

In the presence of an external potential, when the Bohm potential vanishes, we can use Eq. \eqref{fW} to construct a solution to the quantum problem. We show how to proceed for standard potentials.

\subsection{Simple solution for the quantum harmonic oscillator}

The best known solutions for the wavefunctions of a quantum harmonic oscillator (written in terms of Hermite polynomials) have non--vanishing Bohm potentials.

One usual solution to the Schr\"odinger equation for the harmonic oscillator potential $V=m \omega^2 x^2/2$ (with constant frequency $\omega$) can be recovered from the solution \eqref{fW} for vanishing Bohm potential, when $a(t)=0$, and 
\begin{eqnarray}
b(t)=\sqrt{\frac{\alpha}{\sin\left(\omega(t-t_i) \right)}}\, ,\quad c(t)=-\frac{\alpha x_i}{\tan\left(\omega(t-t_i) \right)}\, , 
\end{eqnarray}
for constant $\alpha$ and $t_i$.
This solution allows us to find the amplitude and phase of the wavefunction
\begin{eqnarray}
A(x,t)&=& \sqrt{\frac{\alpha}{\sin\left(\omega(t-t_i) \right)}}\, ,\nonumber\\
S(x,t)&=&\frac{m \omega\left(x^2+x_i^2\right)}{2\tan\left(\omega(t-t_i) \right)}-\frac{m\omega x x_i}{\sin\left(\omega(t-t_i) \right)}\, .
\label{harmonoscill}
\end{eqnarray}
These are amplitude and the phase for the one standard wavefunction for a particle subject to a harmonic oscillator potential \cite{feynam}. In this case, the amplitude coincides with the one prescribed by the VVM expression \eqref{condVVM}, when taking $\alpha=i m\omega/(2\pi\hbar)$.

\subsection{A different solution for the harmonic oscillator}
\label{anothersolharosc1}

It is not trivial to show that a particle subject to the harmonic oscillator potential $V=m\omega^2 x^2/2$
can have another solution completely different to the previous one, with vanishing Bohm potential. In Eq.~\eqref{fW}, let us take $a(t)=0=c(t)$, and
\begin{equation}
    b(t)={\cos^{-1/2}\left(\omega(t-t_i)\right)} \, .
\end{equation}
Thereby, we can calculate
\begin{eqnarray}\label{harmonoscill2}
A(x,t)&=& \cos^{-1/2}\left(\omega(t-t_i)\right)\, ,\nonumber\\
S(x,t)&=&-\frac{m \omega x^2}{2}\tan\left(\omega(t-t_i) \right)\, .
\end{eqnarray}

It can be straightforwardly proved that this is an exact solution for the harmonic oscillator, with vanishing Bohm potential. Solution \eqref{harmonoscill2} is different from \eqref{harmonoscill}, and the exact amplitude is {\it {not}} described by the VVM expression \eqref{condVVM}.

\subsection{Another different solution  for the harmonic oscillator}
\label{anothersolharosc2}

Another solution can be found for the harmonic oscillator by taking, in solution
\eqref{fW}, the conditions
$b(t)=0=c(t)$, and
\begin{equation}
    a(t)={\cos^{-3/2}\left(\omega(t-t_i)\right)} \, .
\end{equation}
For this case, we get
\begin{eqnarray}\label{harmonoscill3}
A(x,t)&=& x\cos^{-3/2}\left(\omega(t-t_i)\right)\, ,\nonumber\\
S(x,t)&=&-\frac{m \omega x^2}{2}\tan\left(\omega(t-t_i) \right)\, .
\end{eqnarray}
This is also an exact solution for the harmonic oscillator with vanishing Bohm potential. The exact amplitude is {\it {not}} given by the VVM expression \eqref{condVVM}.

Notice that solutions \ref{anothersolharosc1} and \ref{anothersolharosc2} have the same phase, but different amplitudes, and therefore those two packets can be differentiated from each other.

\subsection{Time--independent forces}

A whole family of time--independent forces can be proved to solve Schr\"odinger equation with vanishing Bohm potential.
In Eq.~\eqref{fW}, let us consider  the case
\begin{equation}
a(t)=a\ e^{-\mu t}\ ,\quad b(t)=b\ e^{-\mu t}\ ,\quad  c(t)=c\ e^{-2 \mu t}\ ,  \label{ex1}
\end{equation}
where $a$, $b$, $c$, and $\mu$ are constants. These choices produce time--independent forces \eqref{newschFW}. Let us point out that the zero force case as well as (a repulsive) Hooke's force are included among them.

We can analyze few special cases. Take $a=c=0$ to get
\begin{equation}
F= 4 m \mu^2 x \ ,\label{ac}  
\end{equation}
while for $b=c=0$, the force is
\begin{equation}
F = \frac{4}{9} m \mu^2 x  \ .\label{bc}  
\end{equation}
Both cases are (repulsive) Hooke forces.
If we consider only $a=0$, we obtain
\begin{equation}
F = 4 m(\frac{c}{b^2} + x)\mu^2\, , \label{a}    
\end{equation}
a (repulsive) Hooke force plus a constant force. On the other hand, for $b=0$, we get
\begin{equation}
F=\frac{4 m \mu^2 (-18 a c^2 - 3 a^3 c x^3 + a^5 x^6)}{9 a^5 x^5} \, , \label{b}   
\end{equation}
a (repulsive) Hooke force plus (or minus) a centrifugal barrier force plus a force proportional to $x^{-5}$.

Finally, when $\mu=0$, one gets a free particle with $F = 0$.

\section{Particles under non--zero external potential with non--zero Bohm potential}

Several different potentials can be found in a straightforward  form as exact solution
for systems with non--vanishing Bohm potential. Below we show families of such potentials that allows to solve Schr\"odinger equation in an exact manner

\subsection{Harmonic oscillator and $1/x^2$ potential}
\label{harmi1x2sol}

Consider
\begin{equation}
    f(x,t)=x^n \cos^{-n}(\omega (t-t_i))\, ,
\end{equation}
for a constant $n$, and frequency $\omega$. For this case
\begin{eqnarray}
A(x,t)&=&\sqrt{n\, x^{n-1}\cos^{-n}(\omega (t-t_i))}\, ,\nonumber\\
S(x,t)&=& -\frac{m\omega x^2}{2}\tan(\omega (t-t_i))\, .
\label{phaseamplihar1r2}
\end{eqnarray}
These amplitude and phase correspond to wavefunction
with a Bohm potential
\begin{equation}
    V_B(x,t)=-\frac{\hbar^2(n-1)(n-3)}{8m x^2}\, .
\end{equation}
which vanishes only in the cases $n=1$ and $n=3$. Notice that those two cases correspond to the ones studied in Secs. \ref{anothersolharosc1} and \ref{anothersolharosc2}.
Also notice that this solution have the same phase than those in Secs. \ref{anothersolharosc1} and \ref{anothersolharosc2}, but different amplitude.

In this way, solution
\eqref{phaseamplihar1r2} solves equation \eqref{newschV} for the total potential $V(x,t)+V_B(x,t)=m\omega^2 x^2/2$ of the  harmonic oscillator, and thus, for the external potential
\begin{eqnarray}
V(x,t)=\frac{1}{2}m\omega^2 x^2-\frac{\hbar^2(n-1)(n-3)}{8m x^2}\, .
\end{eqnarray}
Therefore, any harmonic oscillator minus a quantum potential $x^{-2}$ \cite{essin} can be solved by the presented solution.

\subsection{Position--independent forces}

In this case, we are looking for a solution for a potential $V(x,t)=-F(t) x$, with a time--dependent force $F(t)=-V'$. 
Let us consider
\begin{eqnarray}
f(x,t)=\int \left[{\mbox{G}}\left(\frac{\beta}{\hbar^{2/3}} x+\zeta(t)\right)\right]^2dx\, ,
\label{ansatzGen1}
\end{eqnarray}
for arbitrary functions ${\mbox{G}}$ and $\zeta(t)$, and constant $\beta$. Thereby, the amplitude is in terms of an arbitrary function, while the phase is
\begin{eqnarray}
A(x,t)&=&{\mbox{G}}\left(\frac{\beta}{\hbar^{2/3}} x+\zeta(t)\right)\, ,\nonumber\\
S(x,t)&=& -\frac{\hbar^{2/3} m }{\beta}x\, \dot\zeta(t)+\mu(t)\, .
\label{ansatzGen2}
\end{eqnarray}
In order to solve Eq.~\eqref{newschV} for the potential $V(x,t)=-F(t) x$, we need to choose that
\begin{equation}
    \frac{d^2{\mbox{G}}(y)}{dy^2}=\pm y\,  {\mbox{G}}(y)\, ,
    \label{airypropert}
\end{equation}
for the argument $y\equiv {\beta}x/{\hbar}^{2/3}+\zeta(t)$ of the function ${\mbox{G}}$.
This implies that this function is an Airy function. 
In this way, this solution has clearly a non--zero Bohm potential
\begin{equation}
    V_B(x,t)=\mp\frac{\hbar^{2/3} \beta^2}{2m}\left(\frac{\beta}{\hbar^{2/3}} x+\zeta(t)\right)\, ,
\end{equation}
whose time dependence is through $\zeta$.
With all these conditions, 
and by chosing
\begin{eqnarray}
\mu(t)=\int\left(-\frac{\hbar^{4/3}m}{2\beta^2}\dot\zeta^2\pm\frac{\hbar^{2/3}\beta^2}{2m}\zeta \right) dt\, ,
\end{eqnarray}
Eq.~\eqref{newschV} is solved for the force
\begin{equation}
    F(t)=\frac{\hbar^{2/3}m}{\beta}\ddot\zeta(t)\pm \frac{\beta^3}{2m}\, ,
\end{equation}
which is determined by $\zeta$.

Notice that this wave solution is always accelerating or decelerating, independent of $\zeta$, with constant acceleration or deceleration (depending the chosen solution) given by the spatial derivative of Bohm potential
\begin{equation}
    -\frac{V_B'}{m}=\pm\frac{\beta^3}{2m^2}\, .
\end{equation}
In this way, any solution
\eqref{ansatzGen2} with property 
\eqref{airypropert} produce accelerating or decelerating wave packets under position--independent forces.

In vacuum, $F(t)=0$,  
$\zeta(t)=\mp\beta^4 t^2/4 m^2\hbar^{2/3} $, and $\mu=- \beta^6 t^3/12 m^3$, thus
recovering the accelerating Airy wave packet  of Sec.~\ref{airysectionberry}, appropriately choosing the upper sign solution \cite{berry}.

\subsection{Attractive or repulsive harmonic oscillators with non--vanishing Bohm potential}

In Sec.~\ref{harmi1x2sol} we obtain a solution for harmonic oscillator potential that require the appearance of a $1/x^2$
potential. In this section we show a wave solution for a pure attractive or repulsive harmonic oscillator, that presents non--constant acceleration.

Let us again start from the solutions
\eqref{ansatzGen1} and \eqref{ansatzGen2}. However, now let us choose the arbitrary function ${\mbox{G}}$ satisfying
\begin{equation}
    \frac{d^2{\mbox{G}}(y)}{dy^2}=\pm y^2\,  {\mbox{G}}(y)\, .
\end{equation}
Thus, the function ${\mbox{G}}$ are Weber functions.  
The Bohm potential is now given by
\begin{equation}
    V_B(x,t)=\mp\frac{\hbar^{2/3} \beta^2}{2m}\left(\frac{\beta}{\hbar^{2/3}} x+\zeta(t)\right)^2\, ,
\end{equation}
Eq.~\eqref{newschV} is solved for the attractive $(+)$ or repulsive $(-)$ harmonic oscillator potentials
\begin{eqnarray}
V(x)=\pm\frac{1}{2}m\omega^2 x^2\, ,
\end{eqnarray}
for a frequency $\omega=\beta^2/m h^{1/3}$, with the  function $\zeta(t)$ fulfilling
\begin{equation}
    \ddot\zeta\pm \omega^2 \zeta=0\, ,
\end{equation}
and 
\begin{eqnarray}
\mu(t)=\int\left(-\frac{\hbar}{2\omega}\dot\zeta^2\pm\frac{\hbar\omega}{2}\zeta^2 \right) dt\, .
\end{eqnarray}
In general, the above solution has the time--dependent acceleration (or deceleration)
\begin{equation}
    -\frac{1}{m}\left(V'+V_B'\right)=\pm\frac{\beta^3}{m^2}\zeta(t) \, .
\label{bohpotenhasnovmasomenos}
\end{equation}

The harmonic oscillator is obtained by choosing the upper sign solution with $\zeta(t)=\zeta_0 \cos(\omega t)$. In this case, we are describing a wave packet that has a non--constant acceleration produced by force \eqref{bohpotenhasnovmasomenos}, explicitly given as
$\zeta_0 \beta^3 \cos(\omega t)/m^2$. This wave packet oscillates.

On the other hand, the repulsive harmonic oscillator is obtained with the lower sign solution,
when $\zeta(t)=\zeta_0 \exp(-\omega t)$. The wave packet experiences a deceleration given by
$-\zeta_0 \beta^3 \exp(-\omega t)/m^2$.
With this, we have generalized the results presented in \cite{yuce}.

\subsection{General solution with non--vanishing Bohm potential}

We can explore a general solution using the solutions
\eqref{ansatzGen1} and \eqref{ansatzGen2}, and an arbitrary function satisfying
\begin{equation}
    \frac{d^2{\mbox{G}}(y)}{dy^2}=\pm y^n\,  {\mbox{G}}(y)\, .
\end{equation}
for integer $n>0$. The Bohm potential in this case, that again produces non--constant acceleration, is
\begin{equation}
    V_B(x,t)=\mp\frac{\hbar^{2/3} \beta^2}{2m}\left(\frac{\beta}{\hbar^{2/3}} x+\zeta(t)\right)^n\, .
\end{equation}

By choosing
\begin{eqnarray}
\mu(t)=\int\left(-\frac{\hbar^{4/3}m}{2\beta^2}\dot\zeta^2\pm\frac{\hbar^{2/3}\beta^2}{2m}\zeta^n \right) dt\, ,
\end{eqnarray}
Eq.~\eqref{newschV} is solved for any potential with the form
\begin{eqnarray}
    V(x,t)&=&\pm\frac{\hbar^{2/3}\beta^2}{2m}\left[\left(\frac{\beta}{\hbar^{2/3}}x+\zeta(t)\right)^n-\zeta(t)^n \right]\nonumber\\
    &&+\frac{\hbar^{2/3}m}{\beta} x\, \ddot \zeta(t)\, .
    \label{potencialgeneral}
\end{eqnarray}
General potentials with the polynomial form \eqref{potencialgeneral} can be solved through accelerating wave packets. The specific form of the potential is given by $\zeta(t)$ which can be choosen freely.
For this solution, it is not possible to construct a time--independent potential for $n>2$.

Finally, all these  wave packets experience a general time--dependent acceleration given by
\begin{equation}
    -\frac{1}{m}\left(V'+V_B'\right)=-\frac{\hbar^{2/3}}{\beta}\ddot\zeta(t) \, ,
\label{bohpotenhasnovmasomenos2}
\end{equation}
and therefore, the functionality of $\zeta$ determines the evolution of the wave packet.

\section{Discussion and Outlook}

We have shown that our approach allows us to find exact solutions to the Schr\"odinger equations for several different external potentials, with either vanishing or non--vanishing Bohm potentials for both free and interacting particles. The main procedure described by Eq.~\eqref{newschV} for an one--dimensional configuration can be extended to two-- and three--dimensional systems \cite{Makowski,sahfaz20201}.

Several of our solutions are, to the best of our knowledge, new or generalizations of previous known ones. However, one important result of this work is the realization on how  a non--vanishing Bohm potential has a non--trivial impact on the evolution of the wavefunctions. This can be seen in the case of free particle solutions, for instance, or in accelerating wavepackets solutions, where the Bohm potential is, at least, partially responsible for producing the acceleration.

It should  be also emphasized that the Bohm potential plays a remarkable role in any wave equation irrespective of its classical or quantum character, producing non--geodesic wave propagation in vacuum as well as on the presence of gravitational fields, birrefringence in anisotropic spacetimes and coupling of polarization with rotation of the gravitational  backgrounds, among other unexpected effects  (see, for instance \cite{sahfaz20201} and references therein).

We are currently carrying out research to extend the results presented here to multiple dimensions, to create new applications of wavefunctions with non--vanishing Bohm potentials and to construct quantum propagators using the techniques developed here.

%%%%%%%%%%%%

\end{document}